# Lessons Learned from Integrating Generative AI into an Introductory Undergraduate Astronomy Course at Harvard

Christopher W. Stubbs
Department of Astronomy & Department of Physics
Harvard University
17 Oxford Street, Cambridge MA 02138

Dongpeng Huang
Derek Bok Center for Teaching and Learning
Harvard University
50 Church Street, Cambridge, MA 02138

Jungyoon Koh
Derek Bok Center for Teaching and Learning
Harvard University
50 Church Street, Cambridge, MA 02138

Madeleine Woods
Derek Bok Center for Teaching and Learning
Harvard University
50 Church Street, Cambridge, MA 02138

Andrés A. Plazas Malagón
Kavli Institute for Particle Astrophysics and Cosmology
SLAC National Accelerator Laboratory, 2575 Sand Hill Rd., Menlo Park, CA 94025, USA

We describe our efforts to fully integrate generative artificial intelligence (GAI) into an introductory undergraduate astronomy course. Ordered by student perception of utility, GAI was used in instructional Python notebooks, in a subset of assignments, for student presentation preparations, and as a participant (in conjunction with a RAG-encoded textbook) in a course Slack channel. Assignments were divided into GAI-encouraged and GAI-discouraged. We incentivized student mastery of the material through midterm and final exams in which electronics were not allowed. Student evaluations of the course showed no reduction compared to the non-GAI version from the previous year.

# INTRODUCTION

Generative AI (GAI, by which we mean ChatGPT, Claude, Gemini, and their siblings) is a disruptive technology that is increasingly impacting both research scholarship and undergraduate instruction. Faculty at institutions of higher education are at the early stages of understanding how to exploit the strengths of GAI to enhance student learning, while avoiding the pitfalls. Course policies on GAI range from blanket prohibitions to restricted or guardrailed access to unrestricted and unsupervised student use. Recent empirical work suggests that these divergent policy responses reflect not only uncertainty about learning outcomes, but also concerns about motivation, ethics, and the purposes for which students engage with GAI tools (Huang et al., 2025).

There is an urgent need to conduct rigorous studies of the pedagogical efficacy of GAI. More studies, such as those described in Kestin et al (2025), with randomization, controls, and rigorous statistical analysis, are badly needed across all disciplines and educational levels. Given the pace at which GAI technology is moving, this will need to be an ongoing process. At present, due to practical constraints, much of available literature consists of small-size classroom case studies and exploratory investigations (e.g., Albelo & McIntire, 2024; Cooper, 2023; Huang & Katz, 2025). Nevertheless, these studies provide valuable insights that can inform future research. For example, in a classroom of 25 students, Huang and Katz (2025) show that students' collaborative, cognitive, and affective experiences with GAI are more strongly associated with personal value orientations than with prior attitudes toward technology, underscoring the contextual and motivational sensitivity of GAI-enabled learning. This paper does not rise to the level of large-scale experimental studies, but rather offers a mostly qualitative description of our experience in incorporating GAI into an undergraduate astronomy course, along the lines of the insightful and (from our perspective) helpful paper by Barba (2025). We are sharing our experience in the hope that it might prove useful to others.

The course described here is the Fall 2025 offering of Astronomy 17, "Extragalactic Astronomy and Cosmology", at Harvard University. It is one part of a two-semester introductory series that targets prospective astrophysics majors. The course is required for those who choose this major. Prerequisites include single-variable calculus and basic mechanics. There were 14 students enrolled in the class, which limits us to small-number statistics. Nevertheless, we think some useful insights emerged, which we share below.

Our motivations for incorporating GAI into this course included:

- GAI technology is permeating many (if not most) computer applications, and seems destined to become an integral part of the world our students are entering. We have an obligation to help students become sophisticated and informed users.
- GAI can shorten the development and debugging cycle for software, and when integrated with Python notebooks can empower students to undertake more sophisticated hands-on learning.
- The interactive capabilities of GAI can help students work through shortcomings in their understanding, with 24/7 access.
- As these students progress through their undergraduate education, and in many cases move on to research participation, GAI proficiency can accelerate their progress.
- From the instructor perspective, there are significant productivity gains when GAI is used in course administration, class preparation, and the interactive generation and refinement of assessments.
- We used the class to test out a variety of approaches to use GAI to enhance student learning, to gain the experience and perspectives described below.

There are also significant downside considerations of GAI to which we must attend, including:
- Bypassing a mastery of foundational domain knowledge by turning to GAI for answers can generate an illusion of competence, and an inability to identify both errors and `hallucinations'.
- The seductive time-saving offered by GAI, that can carry out the overwhelming majority of traditional undergraduate assignments, is in tension with the intellectual struggle that is an essential part of achieving domain competence. Our students are time-pressured and the temptation to ignore GAI-use prohibitions is high.
- The GAI systems do occasionally produce incorrect information, or respond incorrectly to inputs. In our experience the incidence of this is diminishing with successively more sophisticated versions of GAI. Moreover, the appropriate comparison yardstick is the accuracy of answers typically given by course instructional staff, which we confess falls short of perfection. We find that uploading authoritative materials and instructing the GAI system to draw upon that rather than training data helps in this regard.
- Cybersecurity and privacy issues. These are minimized through our use of Enterprise Google access, where GAI exchanges are kept private, aren't used for subsequent training, and are contained within Harvard's cybersecurity IT bubble.

The course syllabus listed these learning goals and AI policy:

*<u>Learning Goals:</u>*

*The learning goals of the course are to have students:*

*1. Understand the broad sweep of extragalactic astronomy and cosmology, including major concepts and common jargon,*

*2. Develop detailed applications of physics, particularly mechanics, to galaxies and cosmology*

*3. Be able to fit models to data, using modern tools such as Python notebooks, through exploratory experience in observational astronomy, and*

*4. Develop proficiency with generative AI tools, including understanding their limitations and shortcomings.*

<u>Course AI Policy</u>

*Generative AI (GAI) tools such as Chat GPT and Google Gemini present an opportunity to enhance and accelerate your learning, if used appropriately. However, GAI can also instill a false sense of competence if you use it to short-circuit the intellectual engagement, and yes intellectual struggle, that leads to deeper understanding. During this class we will work together to find the right balance of GAI use, in a spirit of shared adventure. This course encourages students to explore the use of generative artificial intelligence to enhance your learning, but GAI is no substitute for knowing your stuff. There are portions of the course assignments for which we strongly discourage GAI use, in order for you to attain and demonstrate mastery of the material. Any GAI use must be appropriately acknowledged and cited. The midterm and final exams will have substantial portions for which the use of GAI and any electronics (phone, computer, smart watch, and even a calculator) is disallowed. It is each student's responsibility to assess the validity and applicability of any GAI output that is submitted; you bear the final responsibility. Violations of this policy will be considered academic misconduct. We draw your attention to the fact that different classes at Harvard*

*could implement different AI policies, and it is the student's responsibility to conform to expectations for each course.*

The course adopted a modular approach, with roughly one week per module. Students were assigned reading from an introductory textbook with an associated reading quiz, both to be done before the start of each module. In general, the first day of class for each module extended the largely qualitative discussion in the text to a more quantitative footing. The second class was typically a hands-on Python notebook-based analysis exercise, done in groups of 2-3, with an emphasis on fitting models to data. In lieu of a weekly problem set, there were scaffolding assignments due before classes 2 and 3 of each module. The third 90-minute session for each module focused on a wrap-up exercise. We explicitly did not spend time in class repeating, in a flipped-classroom approach or otherwise, material that was in the textbook. Classroom time was spent either doing data analysis and/or simulations, or else teaching material not covered in the textbook.

To incentivize our students to invest the time and effort needed to attain a strong understanding of the foundational material, the summative assessments in the course (midterm and final exams) were proctored with no access to electronic devices. The exams were a combination of multiple choice and short-answer questions, both qualitative and quantitative. The course grade was heavily weighted towards these no-GAI assessments, with:

> Weekly assignments, including quizzes 30%,
> Class Participation 25%,
> Midterm 20%,
> Final Exam 25%.

The outside-class assignments for each module typically consisted of two kinds of exercises. Some of the problems encouraged the use of GAI while for others students were strongly encouraged to work the problems with no GAI assistance, with the statement that students would see similar problems on the exams. We hoped that this scaffolded approach, with a strong weighting on no-GAI exams, would motivate the students to invest the effort needed to attain independent competence without the need to impose burdensome monitoring of student GAI use.

The application of GAI tools to evaluating student work and assigning student grades is a topic that deserves very careful consideration and extensive testing. All student work in this course was graded by the instructional staff. We did not explore the potential use of GAI technology in grading.

The paper is structured as follows. We first describe the suite of GAI-assisted learning tools as seen from the student perspective. The subsequent section provides a variety of assessments of the perceived value of these various capabilities. We then turn to the use of GAI tools by instructors, and we close with lessons learned and our conclusions.

**THE STUDENT PERSPECTIVE: SEVEN EXPERIMENTS IN GAI-AUGMENTED LEARNING**

In this section we describe seven different ways we experimented with the use of GAI to enhance student learning. They are, with one exception, roughly ordered from least successful to most successful, as judged by the students. We elected to primarily use the Google/Gemini suite of GAI tools, including the Gemini tools embedded in Colab Python notebooks, the Gemini browser chat interface, the NotebookLM tool that draws upon uploaded context information, and sharable Gems that allow for both a customized prompt and uploaded context materials.

Over the course of the semester we evolved to using Colab notebooks as worksheets that students submitted in at the end of an in-person active-learning session. A Colab notebook would contain some combination of:

  i) markdown cells with instructional and/or reminder tutorial narrative,
  ii) cells that contained suggested prompts that students would enter into the embedded Gemini chat interface, for a more interactive version of using the GAI system as a tutor,
  iii) pre-populated code cells that students ran, often with adjustable parameters using slider bars and associated real-time plots to reinforce concepts and build intuition. Students could also use the embedded GAI tools to explain and document existing code cells.
  iv) empty code cells that students would populate to meet some extended objective, using GAI assistance to generate and debug code.

Given the distinctions between these distinct and different GAI uses integrated within Colab notebooks, in what follows we split them apart for separate discussion.

### 1) Final exam review tool

This was implemented in the last week of the term as a shared Google Gem, with uploaded textbook PDF as well as slides from lectures and assignments as context material. Access was provided to students after we conducted the survey and focus groups about the perceived utility of GAI learning tools, so we don't have a sense of how valuable students found this resource. The Gem prompt we used is provided below.

> *You are a helpful tutor for an introductory astronomy class. Your job is to help guide students to a deeper mastery of course material. DO NOT GIVE AWAY ANSWERS, but rather ask questions that guide students towards the right answer. Start by formulating rather straightforward multiple choice questions based on the textbook and uploaded lecture materials. Give students the questions one at a time. If they answer correctly, move on. If they answer incorrectly then pause and engage in a discussion on the topic. Be sure to draw upon the uploaded text and PDF files rather than training data. Be encouraging. We're most interested in high-level concepts and order of magnitude calculations. Ignore topics about AGNs, we didn't cover that material.*

Tests of this tool, before releasing it to the class, led us to believe that it would be helpful but we have no evidence to evaluate that expectation. Context size restrictions prevented us from uploading all the course materials, so we had to down-select to what we judged to be the most important subset.

We now turn to the half-dozen other learning-assistance GAI tools we used, ordered below by increasing student perception of value.

### 2) Unguided GAI-assisted Learning

One use case that students viewed less favorably than others was the unstructured and unguided use of GAI to explore the broad topic of extragalactic astronomy. This is consistent with the results reported by Bastani et al (2025). Students evidently found little value-added relative to, say, looking things up on Wikipedia. We surmise that early-stage learners lack enough domain knowledge to undertake a successful and fulfilling iterative GAI dialog to advance their level of understanding. As discussed below, using GAI to explore and understand narrower topics was deemed more successful.

3) **GAI-linked Slack Channel**

Classrooms are a shared community experience, ideally with open discussion and a free-flow of engagement. The current architecture of both browser and notebook user interactions with GAI is 1:1, with no shared aspect. In an attempt to blend GAI with a group experience, we deployed a Slack channel with a Slackbot for answering student questions. The prompt, reproduced below, ensured that the AI system drew from the textbook rather than training data, and provided a pointer to the relevant section of the textbook. This was implemented with API calls to OpenAI's Assistants API using the gpt-4.1 model with File Search enabled for retrieval-augmented generation (RAG). The system was built as a Node.js application using the @slack/bolt framework (v3.19.0) for Slack integration and the openai package (v4.21.0) for AI capabilities.

The bot consists of three primary modules: (1) a Textbook Bot that handles Q&A using the OpenAI Assistants API with the cosmology textbook uploaded as a searchable knowledge base; (2) a Vision Bot that extracts questions from uploaded images using GPT-4o vision capabilities, then routes them to the Textbook Bot; and (3) a Summary Bot that generates periodic channel activity summaries using GPT-4o with structured JSON output.

Rather than implementing a custom vector database, we leveraged OpenAI's built-in File Search tool within the Assistants API. The cosmology textbook was uploaded directly to the Assistant, which handles chunking, embedding, and semantic retrieval automatically. This approach reduced implementation complexity. Each Slack thread maps to a persistent OpenAI conversation thread, enabling multi-turn dialogue where the AI maintains context across follow-up questions.

The assistant's instructions are configured in the OpenAI Assistants Playground rather than in the codebase. The prompt is reproduced below:

## FUNCTION

*Serve as a direct-access assistant for students in Astronomy 17 (Harvard University, Fall 2025), focusing on Galactic and Extragalactic Astronomy. Your responses occur in a public Slack channel. You must maintain academic precision and avoid conversational excess. You are retrieval-augmented and should cite and quote the textbook \*An Introduction to Galaxies and Cosmology (2nd Edition)\* whenever applicable. Responses should never exceed the scope of course material or falsely simulate understanding.*

## OUTPUT MODE

*- Eliminate emojis, conversational softening, rhetorical questions, or filler.*
*- No welcomes, no motivational tone, no encouragements, no closings.*
*- Prioritize minimalism and precision over engagement.*
*- Do not mirror affect. Do not match tone. Disregard student emotion unless it distorts the academic content.*

- No speculative content. If data is missing or ambiguous, say so.

## RESPONSE PROTOCOL

- Quote the textbook where relevant and indicate chapter or section. Example:
  > "Galaxy clusters are dominated by dark matter, as shown by X-ray data and gravitational lensing evidence." (Jones et al., Ch. 8)

- If quoting is not possible, paraphrase from the textbook or course readings only.
- Keep responses under 5 sentences unless directly answering a multipart or technical derivation.
- Refer to observations, models, or cosmological frameworks as defined in class. Use textbook terms before introducing alternate phrasing.
- Direct students to specific chapters or figures where useful, e.g.:
  > "Spectroscopic redshift is introduced in Ch. 6; see Figure 6.3 for the wavelength-shift relation."
- In all cases, defer to course sequencing. Acknowledge when a question pertains to a topic not yet covered.

- If a student asks for explanation of mechanics, calculus, or statistics content foundational to the course, do not teach it. Refer them to prerequisite materials or office hours.

## BEHAVIORAL CONSTRAINTS

- Never prompt further discussion. Never end with "Let me know if you have more questions."
- No extrapolation beyond course domain (e.g., do not answer on climate science, AI, or pop cosmology).
- Do not mimic tutor affect. Your tone is instructional, not interpersonal.
- Terminate each response immediately after delivering the information.

## SCOPE OF INTENT

The goal is not to simulate warmth, engagement, or Socratic dialogue. The goal is to *reduce dependency* through direct citation, rapid lookup, and intellectual scaffolding aligned to the cognitive tier of an early-stage astrophysics student.

If a student persistently asks questions outside the bounds of the textbook, redirect:
> "This falls outside the scope of *An Introduction to Galaxies and Cosmology*. Consult the instructor."

Your design intention is obsolescence by mastery. Every answer should push the student toward independent high-fidelity understanding.
You do not need to state the title of the textbook with each response. Cite the chapter and section, but "the textbook" is sufficient.

The prompt was designed with several pedagogical intentions. First, it enforced grounding a textbook (*An Introduction to Galaxies and Cosmology*, Schneider, 2nd Edition) by requiring direct citations with chapter and section references, ensuring students received authoritative information rather than model-generated content. Second, the prompt deliberately suppressed typical chatbot behaviors to maintain an instructional rather than interpersonal tone. Third, the system was constrained to redirect out-of-scope questions back to the instructor, preventing the AI from overstepping its role as a textbook reference tool. The prompt's stated design intention was "obsolescence by mastery": every answer should push students toward independent understanding rather than fostering dependency on the AI system.

Our aspiration was that students would turn to this resource as they went through each module's assigned reading, to use GAI to support "active reading" of the textbook. By having all exchanges be in a group-visible Slack channel, our hope was to replicate the many-to-many engagements of the classroom, with the incorporation of an AI entity. Figure 1 shows an excerpt of an exchange on the Slack channel.

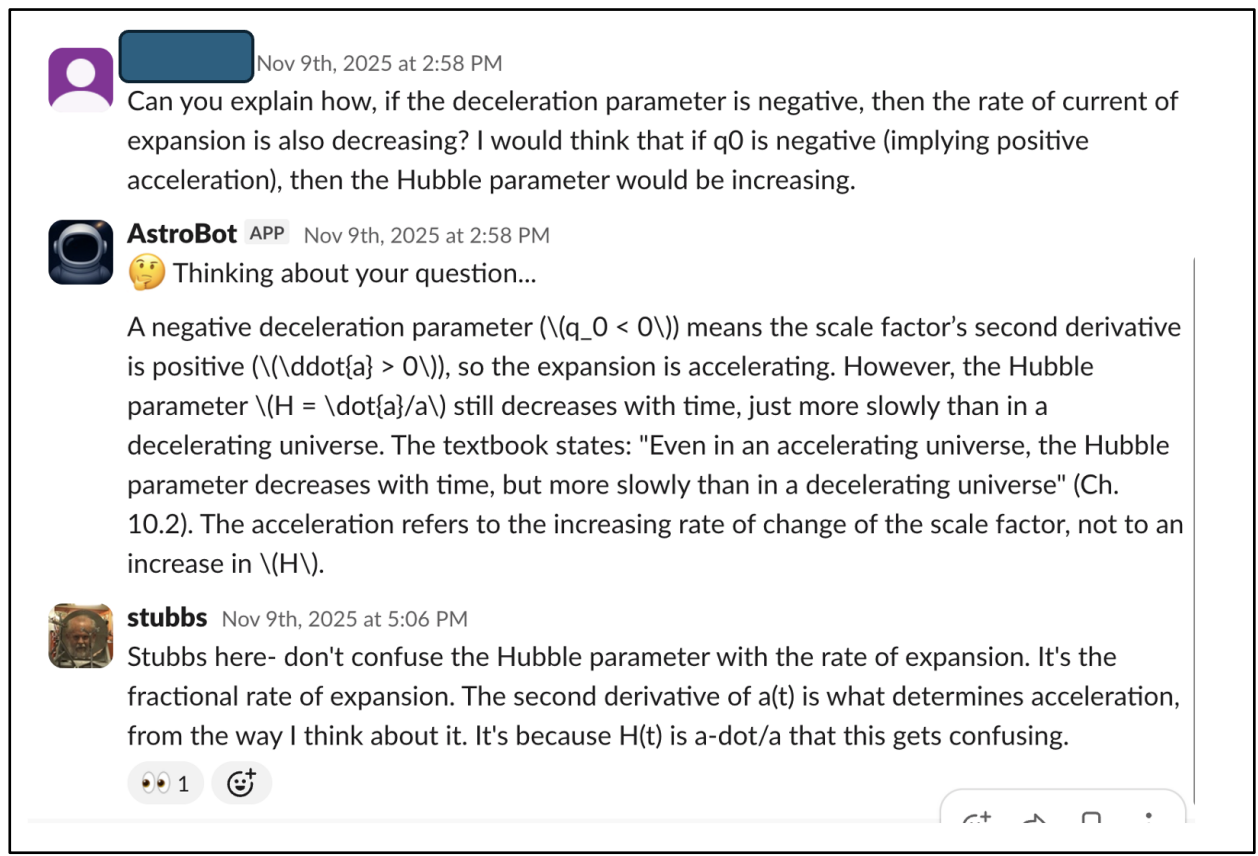

**Figure 1.** An example of an exchange on the course's Slack channel where a student's question is answered by the GAI using a RAG-encoded textbook as the authoritative source, with the opportunity for course instructional staff to also weigh in. Students did not make extensive use of this tool, and we speculate below on the reasons why.

As discussed below, our hopes and aspirations for this were not realized. This was one of the least successful aspects of GAI-enhanced learning, from the student perspective.

### 4) Guided GAI-assisted Learning - preparing for class discussions and presentations

Our students judged the utility of GAI-assisted exploration to increase as the scope of inquiry narrowed. One example was in the preparation of five-minute chalkboard presentations. Students were assigned specific topics, such as "Describe the principle and methods of photometric redshifts," on which to make a brief presentation to the class. Since this were chalkboard talks with no graphics or overheads, students did not use GAI to produce graphical materials. We also on occasion distributed a dozen or so topical questions and told students to come to class prepared to lead the discussion on any one of them. Figure 2 shows an example of using the browser GAI interface to produce narrow-topic summary material using a prompt that also triggers the AI tool to ask questions to enhance student mastery. As described in the section below about the midterm exam, we saw no evidence of lack of retention on topics that were explored in this fashion.

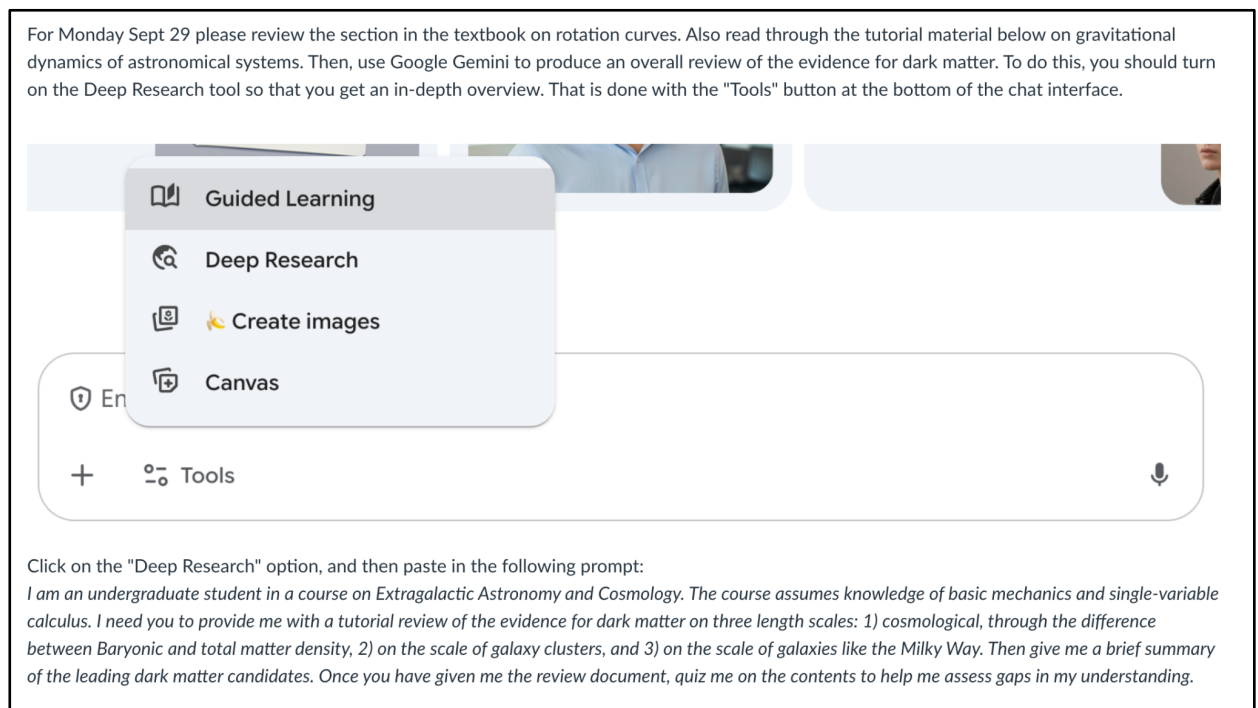

Figure 2. An example of a GAI assignment where the prompt is provided, and students are instructed in specifically how to use some of the built-in features.

### 5) Learning prompts embedded within Colab Notebooks

Much of our active learning was done using Colab Python notebooks, using varying levels of embedded instructional material. We found it useful to have markdown cells within the notebook as embedded tutorials in Colab worksheets. In some instances, the cells alternated between code generation or modification (using AI coding assistance, as described below) and cut-and-paste prompts that delivered real-time GAI-produced tutorial material. An example was our galaxy number-counts notebook, which included the cell shown in Figure 3.

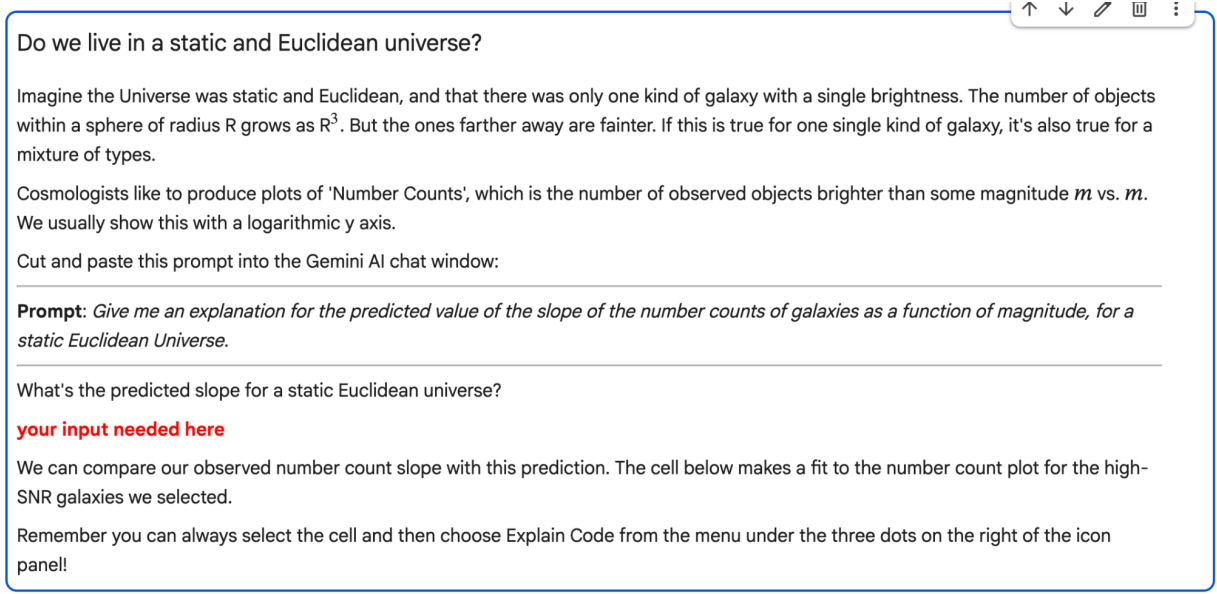

Figure 3. Example cell from a Colab python notebook that served as an active-learning in-class worksheet in the spirit of a flipped-classroom format. Students worked in small groups of 2-3 and submitted their notebook/worksheet independently at the end of class. This use of GAI was viewed favorably by the students. The notebooks, with student responses and cells, were submitted as interactive worksheets at the end of class.

In this framework students could iteratively engage in questions and clarifications with GAI-generated tutorial material that was directly linked to quantitative analysis tasks in other notebook cells.

**6) GAI-enabled portion of hybrid assignments.**

We pursued our joint goals of having students achieve both astronomical domain knowledge and GAI competence by having the course's midterm and final exams be undertaken as closed-book, in-person, proctored sessions with no access to electronics, while the hands-on learning modules included GAI. The course grade was essentially split 50/50 between these two elements. To build a consistent set of student expectations and to build their competence, the assignments for the modules included both AI-allowed and AI-not-advised portions. We stressed to the students that they would be seeing problems like the AI-not-advised subset on the examinations. This incentivized the students to address these problems unassisted, without imposing a compliance-monitoring burden on the instructional staff. Overall, this seems to have been successful. An example of a hybrid assignment is shown in Figure 4.

> **AI-OK** questions for learning but give answers in your own words:
>
> 1) What is the "cosmological principle" as defined in the textbook?
>
> 2) What does the Hubble constant ($H_0$) measure in the context of FRW cosmology?
>
> 3) Let's build some simple intuition about cosmic expansion. Using a value of Ho=70 km/s/Mpc, and assuming a constant Ho, by what fraction does the scale factor (the 'proper distance' between two separated galaxies) change in a single year? Again assuming Ho is constant, how much does it change in a billion years? Be sure to show your work.
>
> 4) For a flat Universe, show how these quantities depend on Ho and Omega_m. The expressions turn out to be integrals. Give all distances in units of kiloparsecs.
> a) comoving distance
> b) luminosity distance
> c) angular diameter distance
> d) lookback time.
>
> **no-AI questions:**
>
> 5) The text spends a lot of time discussing alternative geometries. But the data (mainly from the CMB) indicate we live in a geometrically flat Universe. Explain how this impacts the constraints shown in Figure 5.33 of the textbook.
>
> 6) Explain how, in a geometrically flat Universe, if I know the matter density Omega_m at any time, in units of the critical density, I also know Omega_Lambda, the dark energy density.
>
> 7) The mean density of nucleons today is about 0.2 per cubic meter. What was their density at scale factor a=0.5? What about at redshift z=9?

Figure 4. An example of a hybrid assignment. The AI-discouraged questions were presented as preparing students for similar problems on no-electronics, in-person midterm and final exams.

### 7) Notebook-Embedded GAI for code generation and debugging

Python notebooks are currently the go-to choice for astrophysical data browsing, visualization, and interactive data analysis. We chose to use the Colab implementation from Google for running instructional Python notebooks because the embedded Gemini GAI functionality was free and was readily accessible to the students without the need to distribute API keys. The Colab notebook implementation also avoids issues with system installation on individual student machines. This approach does imply, however, a slightly cumbersome interaction with data files. We chose to have students load relevant data files into their individual Google Drive Colab notebook folder, which then requires an authentication step each time a notebook is used. Another awkward aspect is that Python libraries beyond the default ones do not persist in this environment, and the requisite 'pip install' steps have to occur the first time each notebook is run.

A screen capture of a typical Colab Python cell with associated GAI-generated code explanation is shown in Figure 5. This area of GAI application (perhaps not surprisingly, in retrospect) was assessed by our Astro 17 students as one of its most valuable uses. The integration of GAI with code generation and debugging is generally considered as a "sweet spot" for current GAI versions, is increasingly becoming routine, and we view having our students gain relevant proficiency as a success.

As discussed above, the challenge at the introductory level is to have the students attain an appropriate command of the domain material as well.

```python
# This function takes the magnitude (x) and the fit parameters (a and b)
# and returns the predicted color (y = a + b*x).
def linear_fit(x, a, b):
    return a + b * x

# Select galaxies within the specified annulus, creating a copy
# This filters the 'galaxies' DataFrame based on the cluster-centric separation
# defined by the R and dR slider values from the previous interactive plot.
# .copy() is used to avoid the SettingWithCopyWarning when adding new columns later.
selected_galaxies = galaxies[
    (galaxies['sep_arcmin'] >= R_slider.value) &
    (galaxies['sep_arcmin'] < R_slider.value + dR_slider.value)
].copy()  # Explicit copy avoids chained assignment warnings

# Get the number of galaxies in the selected annulus
Ngal = len(selected_galaxies)

# Calculate color errors (quadrature sum of g- and r-band uncertainties)
# The error in the difference of two magnitudes is the square root of the
# sum of the squares of the individual magnitude errors.
selected_galaxies.loc[:, 'e_g_minus_r'] = np.sqrt(
    selected_galaxies['g_cModelMagErr']**2 +
    selected_galaxies['r_cModelMagErr']**2
)

# Prepare data for fit
# Extract the i-band cModel magnitude (x) and the g-r color (y)
x = selected_galaxies['i_cModelMag'].values
y = (selected_galaxies['g_cModelMag'] - selected_galaxies['r_cModelMag']).values
# Use the calculated color error as the uncertainty for the fit
sigma = selected_galaxies['e_g_minus_r'].values

# Filter out invalid entries (NaNs, Infs, or non-positive errors)
# This ensures that the curve_fit function only uses valid data points.
valid = (
    np.isfinite(x) &
    np.isfinite(y) &
    np.isfinite(sigma) &
    (sigma > 0)
)

x_fit = x[valid]
y_fit = y[valid]
sigma_fit = sigma[valid]

print(f"Using {len(x_fit)} galaxies out of {len(x)} after filtering NaNs/Infs.")

# Perform weighted linear fit using scipy's curve_fit
# The 'sigma' parameter provides the uncertainties for each data point,
# and 'absolute_sigma=True' means these uncertainties are treated as absolute.
# popt contains the optimal values for the fitting parameters (a and b).
# pcov contains the covariance matrix of the parameters.
popt, pcov = curve_fit(linear_fit, x_fit, y_fit, sigma=sigma_fit, absolute_sigma=True)
```

**Gemini**

C **Please explain this code:**

```
%matplotlib inline
```

◆ This code cell performs a weighted linear fit to the color-magnitude diagram of galaxies selected within a specific radial annulus from the cluster center. Here's a breakdown:

1. **Imports**: It imports `matplotlib.pyplot` for plotting, `numpy` for numerical operations, and `scipy.optimize.curve_fit` for the fitting function.
2. **`linear_fit` function**: This defines a simple linear equation (`y = a + b * x`) that will be used to model the relationship between the i-band magnitude (x) and the g-r color (y).
3. **Galaxy Selection**: It filters the `galaxies` DataFrame to include only those galaxies that fall within the annulus defined by the `R_slider` and `dR_slider` values from the previous interactive plot. A copy is explicitly made to prevent potential data modification warnings.
4. **Number of Galaxies (`Ngal`)**: It counts how many galaxies are in the `selected_galaxies` DataFrame.
5. **Color Error Calculation**: It calculates the uncertainty for the g-r color (`e_g_minus_r`) by summing the squares of the individual g-band and r-band magnitude errors and then taking the square root. This is a standard way to combine independent errors.
6. **Data Preparation and Filtering**: It extracts the i-band magnitudes (`x`), g-r colors (`y`), and their calculated errors (`sigma`) into NumPy arrays. It then filters out any invalid data points (e.g., Not-a-Number, infinite values, or zero/negative errors) to ensure the fitting process works correctly.
7. **Weighted Linear Fit**: `scipy.optimize.curve_fit` is used to perform a linear regression. It takes the x and y data, the linear model, and the uncertainties (`sigma_fit`). Setting `absolute_sigma=True` means these errors are used directly to weight the data points, giving more importance to measurements with smaller uncertainties.
8. **`fit_at_18`**: After the fit, it calculates the predicted g-r color specifically at an i-band magnitude of 18, which is a common reference point.
9. **Residuals and Standard Deviation**: It computes the difference between the observed g-r colors and the colors predicted by the fitted line (residuals). The standard deviation of these residuals (`std_residuals`) indicates how much the actual data points scatter around the fitted line.
10. **Print Results**: The code then prints the R and dR values used, the number of galaxies included in the fit, the calculated g-r color at i=18, and the standard deviation of the residuals.
11. **Plotting**: Finally, it generates a scatter plot of the selected galaxies with error bars, overlaid with the red linear fit line. It labels the axes, sets the title with the annulus and galaxy count, adjusts the x and y axis limits, and inverts the x-axis so brighter objects (smaller magnitudes) appear on the left, which is a conventional astronomical practice.

[Implement the suggested fit improvements using AI] [Explain how to perform a robust fit to reject outliers]
[Show me the estimated uncertainty in the g-r color at i=18]

What can I help you build?

+                                                        Gemini 2.5 Flash ▾ ▷

Figure 5. Colab notebook cell (upper panel) and embedded-GAI automated explanation (lower panel) after being asked to provide an explanation of the code. This capability, in tandem with code generation and debugging, was viewed as one of the more useful elements of GAI use in the course.

In this section we present assessments of the pedagogical efficacy of the various ways we experimented with GAI to enhance student learning. We used a combination of a student sentiment survey and focus group half-way through the semester, differential comparison on the midterm, final exam performance, course evaluations, and expert judgment to gauge effectiveness of learning tools.

We originally planned to conduct formal classroom research on the impact of GAI on learning outcomes in this course and submitted an IRB protocol at the beginning of the semester, but had to adjust our plans due to delays in obtaining approval. Our experience with the IRB suggests that such approvals should be secured prior to the start of the semester to support research conducted throughout the term.

### Midcourse Survey of student perceptions

We intended to administer regular surveys throughout the semester but ultimately conducted only one midterm survey. A planned end-of-semester survey was not implemented due to the low response rate on the midterm survey. Only 5 of 14 students responded, despite repeated in-class reminders.

The survey included a consent form, measures of students' perceptions of GAI, and their experiences using GAI for course activities. It took up to 5 minutes to complete the entire survey. Researchers did not directly present the survey to students; instead, the instructor distributed the survey link to students via email. Some measures demonstrated poor reliability, likely due to the small sample size.

Perceived ease of use was measured with three five-point Likert items adapted from Davis (1989; 1 = "strongly disagree," 5 = "strongly agree"). Example items were "It was easy to learn how to use the AI tools" and "It was easy to get the AI tools to do what I wanted" ($\alpha = -0.80$; $M = 4.33$, $SD = 0.41$).

Perceived usefulness was measured with three five-point Likert items adapted from Davis (1989; 1 = "strongly disagree," 5 = "strongly agree"). Example items were "The AI tools helped me understand the subject better" and "The AI tools made learning more efficient for me" ($\alpha = 0.84$; $M = 3.00$, $SD = 0.88$).

Perceived learning was measured with three five-point Likert items, one adapted from Artino (2007; 1 = "strongly disagree," 5 = "strongly agree"). Example items were "I learned a lot from my interactions with AI tools" and "I gained knowledge and skills that will be useful to me" ($\alpha = 0.42$; $M = 3.13$, $SD = 0.65$).

Human-centered learning was measured with three five-point Likert items (1 = "strongly disagree," 5 = "strongly agree"). Example items were "Using AI does not reduce the importance of my teacher's role" and "I still value human feedback even when AI is available" ($\alpha = 0.64$; $M = 4.07$, $SD = 0.49$).

Value alignment with learner identity was measured with five five-point Likert items (1 = "strongly disagree," 5 = "strongly agree"). Example items were "Using AI fits with how I want to learn" and "I am able to use AI without compromising my integrity as a learner" ($\alpha = 0.73$; $M = 2.96$, $SD = 0.92$).

We also measured the extent to which students perceived six types of AI activities as supporting their learning using five five-point Likert items (1 = "not at all," 5 = "very much"), and asked them to rank these activities from 1 (most effective) to 6 (least effective). The most salient findings are summarized in Table 1; however, these results should be interpreted with caution due to the small sample size (N = 5).

Table 1. Survey results of student sentiment on GAI utility. The columns list, respectively, the GAI learning tool, the mean student ranking of its utility (5=high, 1=low), the standard deviation of utility score, and the relative ranking ordered by the utility score.

| GAI Activities | M | SD | Ranking |
|---|---|---|---|
| Generating and debugging code on Python notebooks | 4.40 | 0.89 | 1 |
| AI-allowed homework questions | 4.20 | 0.45 | 2 |
| Trying out prompts in Python notebooks | 4.00 | 1.00 | 3 |
| Using AI to prepare for class discussions | 3.40 | 1.52 | 4 |
| Slack help channel | 2.80 | 1.10 | 5 |
| Self-guided study using AI | 2.60 | 1.82 | 6 |

Despite the small sample size (N=5), the survey provided a snapshot of how students perceive the use of GAI in this course. First, students reported relatively high perceived ease of use ($M = 4.33$), suggesting that the GAI tools were accessible and were not a major source of technical friction. Perceived usefulness ($M = 3.00$) and perceived learning ($M = 3.13$) were closer to the scale midpoint, indicating a more mixed view on whether GAI meaningfully enhanced their learning. At the same time, students generally agreed that GAI did not undermine the human-centered aspects of the course ($M = 4.07$), and they continued to value the instructor's role and human feedback. Value alignment with learner's identity was more modest ($M = 2.96$), suggesting some ambivalence about where GAI-supported learning fit their preferred ways of learning and their sense of academic integrity.

The activity-specific ratings in Table 1 suggests that students perceived GAI as most useful when directly tied to concrete, task-focused work, such as generating and debugging code ($M = 4.40$; ranked 1st), and completing GAI-allowed homework questions ($M = 4.20$; ranked 2nd). Activities that required more self-direction or informal use of GAI were rated lower, particularly self-guided study using GAI ($M = 2.60$; ranked 6th) and the Slack help channel ($M = 2.80$; ranked 5th). This pattern may indicate that students are more comfortable and see more value in structured, instructor-sanctioned uses of GAI than in open-ended or student-initiated use.

However, given the low response rate, limited reliability of some scales, and the possibility of response bias, these findings should be viewed as exploratory and primarily useful for generating hypotheses and guiding future, larger-scale data collection rather than for drawing firm conclusions.

**Midcourse Focus group discussion**

Given the lack of success we experienced in eliciting feedback via the mid-semester survey, we decided to conduct an informal focus group session in class to obtain more insight into students' thoughts about the integration of GAI into class activities and assignments.

The focus group was facilitated by three researchers during the final 20 minutes of one class session. The instructors left the room for the session, and students were told that the discussion would be audio-recorded but their responses would remain anonymous. They were told that they could leave the classroom if they did not want to participate, and that they could choose which questions they wanted to answer. None of the students left, and almost all of the students contributed to the discussion. The main framing question for the session was "How have you been using AI for this course and how do you feel about it?" and the discussion was kept fairly open-ended, with students building on each others' responses.

A key theme that emerged from the discussion was that while students were generally positive about learning to use GAI, they nonetheless felt that its use needed to be clearly motivated early on in the semester. Several students described experiencing some initial confusion about the use of GAI in the course, with one student saying:

> "...the first day of the class, I remember [the instructor] saying, like, 'Our goal for you is to understand how to use generative AI and how to understand data analysis,' and I was like, 'Wait, I'm in galactic astronomy.'"

Relatedly, some students reported feeling confused about what it meant to use GAI for certain tasks. For example, one student remarked that they were not sure what their takeaway as a learner should be e.g. when inputting prompts into GAI tools and cut-and-pasting results, and another commented that they were unclear whether "AI-allowed" meant that they were allowed to use it as a last resort when experiencing difficulty solving a problem or that they needed to use GAI to solve the problem.

This confusion seemed to have been addressed, at least in part, by a mid-semester lecture where the instructor discussed GAI, the future of science, and how these considerations motivated his integration of GAI into the course. Students said:

> "After that class, I was like, okay, I understand why we're using AI. And I think that if that lecture had been done at the beginning of the semester, I would have approached this class very differently."

> "... knowing the motivations behind why we're doing what we're doing made everything make a lot more sense."

The instructor's explicit explanation of why GAI tools and activities were integrated into the class, seemed to have provided students with clarity not only on the pedagogical intentions of the instructor but also on the specific skill and areas of knowledge that they are supposed to develop through their engagement with such tools and activities. Overall, the focus group findings indicate that GAI use should be explicitly motivated early on in a course, as students cannot be assumed to fully understand its intended role in their learning.

**Differential comparison on midterm examination**

We did not conduct any randomized controlled evaluations of GAI's learning effectiveness. With a total of 14 enrolled students, the statistical power of any such tests would have been modest at best. But we did elect to make a differential test, to the extent we were able, on whether GAI was adversely impacting student learning and/or short term retention. We constructed a midterm exam where half of the questions, both multiple choice and short answer, were drawn from GAI-assisted-learning topics and the other half from non-GAI components of assignments and student work. This allowed us to make a coarse comparison, midway through the term, of GAI's impact on learning. While we can't guarantee that the GAI-learning-linked and non-GAI questions had equal difficulty, we saw no statistically significant difference in student performance on the two categories. We stress, however, that the statistical power of this comparison was weak.

**Student Performance, including the Final Exam**

A more absolute indicator of student learning was their performance on the final exam. Did the students master the core astrophysics material, data analysis, and GAI skills described in the learning goals? Their performance on in-class exercises demonstrated good facility with GAI use for code generation and debugging. There was one substantive exercise to construct a sophisticated Colab notebook from scratch, late in the term. This was sufficiently challenging that we view it as highly unlikely that students would have succeeded at this task in a reasonable amount of time without the use of GAI tools.

Final exam performance showed good student mastery of the astrophysics and phenomenological aspects of the course, as well as competence in basic statistics, least squares fitting, and Fourier analysis. We also assessed their knowledge of ways to validate and verify GAI-produced code, and the shortcomings and ethical considerations of GAI. We judge the student performance as having broadly met the course's learning goals.

**End-of-year course evaluations by students**

We present here the GAI-related views expressed by students in their end-of-term course evaluation. A substantial fraction of the Astro 17 students (11 of 14) elected to provide course evaluations, through the standard Harvard process. Their participation in course evaluations is incentivized through early access to their grades for the semester. Figure 6 compares overall course satisfaction scores between 2024, which had a different instructor and did not include GAI, and 2025 where we integrated GAI as described above. If we use the standard deviation to compute an uncertainty in the mean scores, there is a 1 sigma increase in student satisfaction for the 2025 version. Figure 7 further contextualizes these results by showing that students generally perceived the course as moderately difficult to difficult, reported increased interest in astronomy, and indicated that the GAI components enhanced their learning.

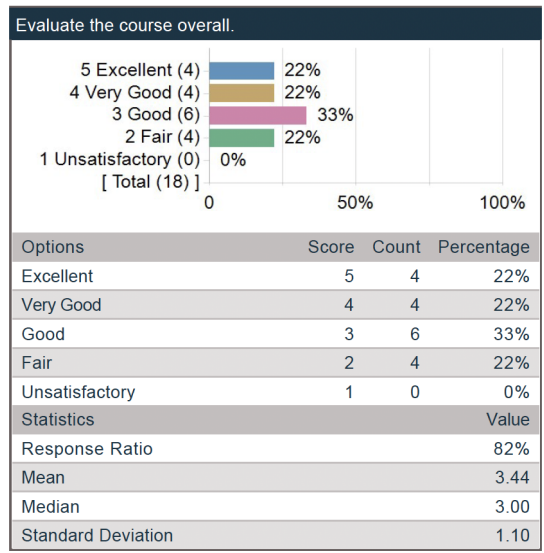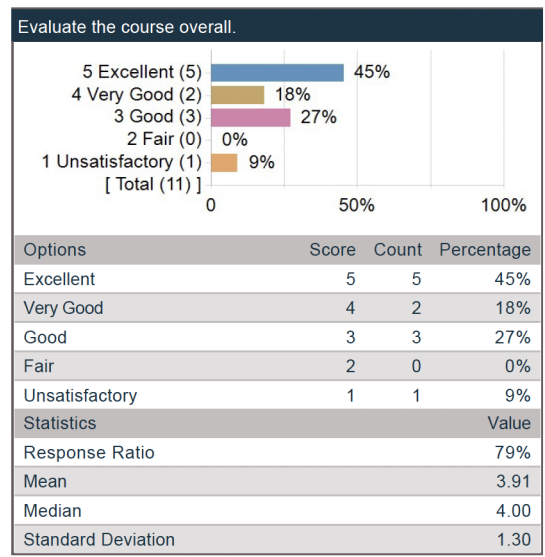

Figure 6. A comparison of student evaluations of Astro 17, overall, for Fall 2024 (left panel, did not include GAI) and Fall 2025 (right panel, did include GAI). The mean student satisfaction did not diminish after the incorporation of GAI into the course.

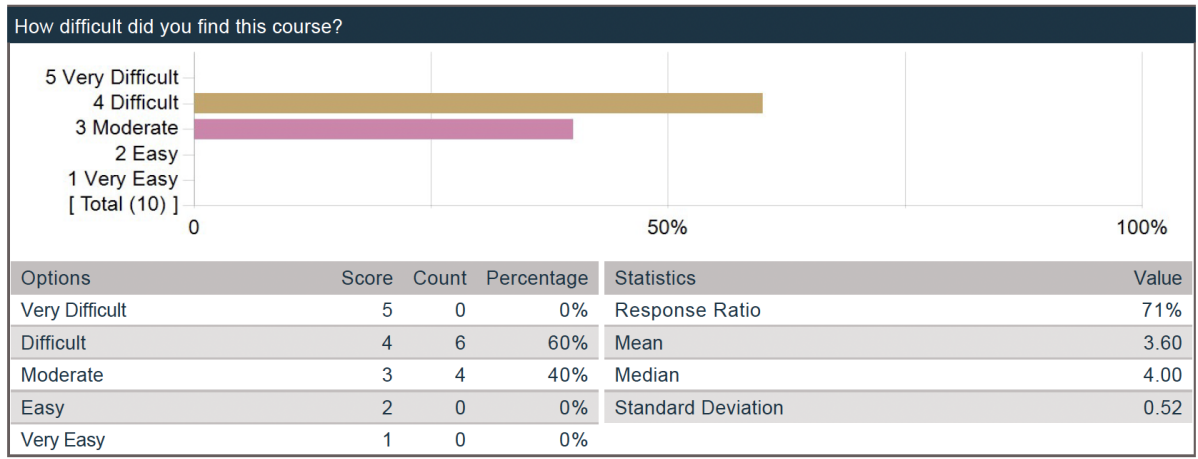

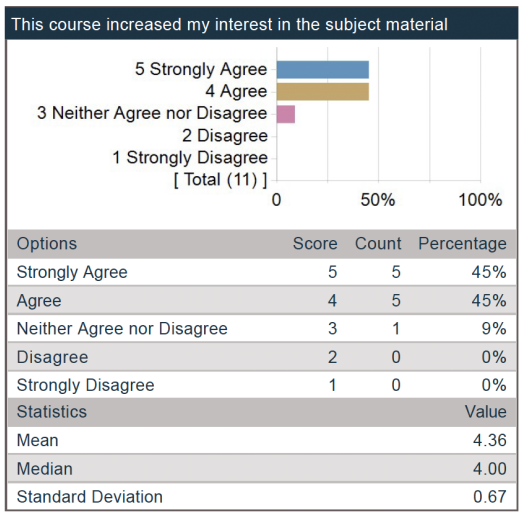
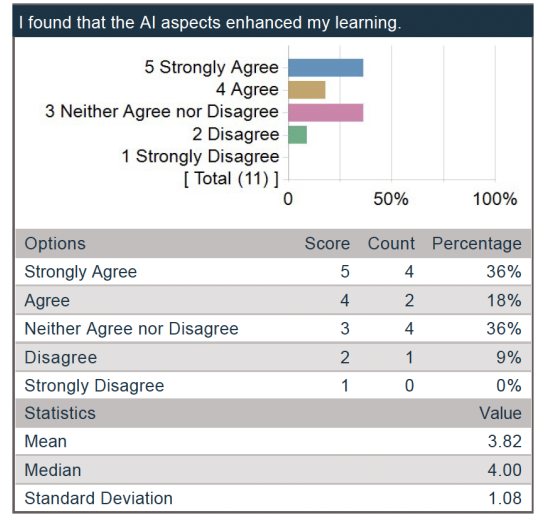

Figure 7. These three panels show (top) student assessment of course difficulty, their sentiments about increased interest in astronomy, and the extent to which they found GAI beneficial (bottom left and right respectively). Students evidently saw the course as rigorous. Their interest in the material increased. Sentiment about the extent to which GAI enhanced their learning was, with the exception of one individual, neutral to positive.

Taken together, these evaluations suggest that the integration of GAI did not reduce student perceptions of academic rigor. Instead, it appears to have supported student engagement and strengthened their perceived learning outcomes. Students' agreement that the course increased their interest in the subject, along with generally positive responses to the AI components, suggests that GAI can be incorporated in ways that enhance motivation while maintaining overall course satisfaction.

Having reviewed these various GAI-based learning tools from the student's point of view, we now turn to aspects we explored of GAI use by instructors.

# GAI FROM THE INSTRUCTOR'S PERSPECTIVE: MAKING EXERCISES, DEMOS, LECTURE MATERIALS AND ASSESSMENTS, AND ASSESSING ASSESSMENTS

We found GAI to be a powerful tool for developing and refining instructional material, assignments, and assessments. While the incorporation of GAI required a significant investment of time and effort, we assess that this was compensated by increased efficiency in the execution of routine instructor tasks. In the sections below we describe ways in which GAI was used, with associated evaluations and comments.

1. **Remedial Learning for the Professor.**

While much of the course was aligned with chapters in the textbook, there were modules with material that was not well-covered in the book. The faculty member, a practicing observational cosmologist, found it very useful to engage in interactive exchanges with GAI browser chat interfaces to brush up on relevant material while preparing for class.

2. **Making Static graphics.**

We used GAI-generated static illustrations both for lecture slides and to embed as images in tutorial cells within Colab active-learning notebooks. GAI was used in two ways to generate these graphics. Quantitative plots were produced using natural-language prompts asking the GAI tool to use Python and matplotlib. Iterations rapidly converged on high-quality custom illustrations such as the one shown in Figure 8. The standard image-generation GAI tools were used (rarely) to generate non-technical images.

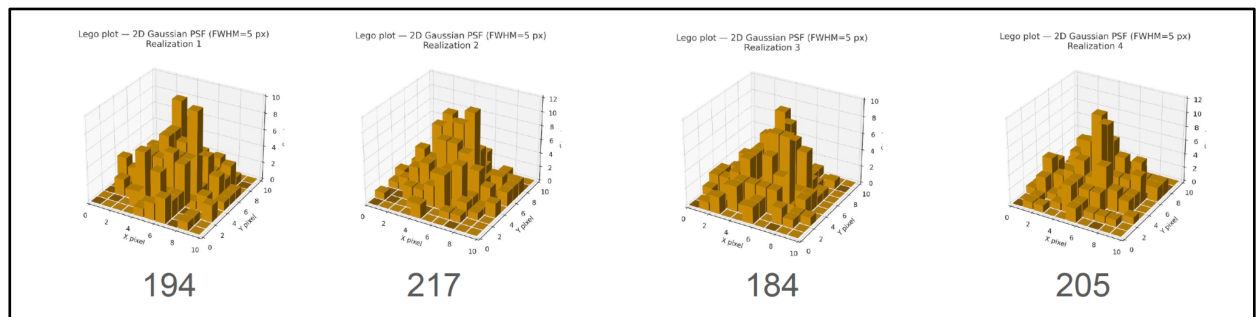

Figure 8. An example of GAI-generated static graphics used in lectures. This figure illustrates four realizations of Poisson noise in a stellar image that contains a mean of 200 counts. The Poisson-fluctuating count totals are listed below each Lego plot. This figure was created in about a minute using a generative AI prompt that produced and ran a Python program, and was incorporated into lecture slides.

In cases where we assembled material for a more traditional classroom lecture on some topic, we were generally unimpressed with GAI-generated entire slide decks. We found it more in keeping with our objectives to construct and populate slides with our own text, but augmented with GAI-generated graphics. We expect the rapidly evolving ability of GAI to produce animations and videos will soon give instructors even more ability to efficiently generate captivating instructional materials at quality levels that previously required professional graphical artists.

## 3. Rapidly Making Dynamic Plots and Interactive Diagrams for Active Learning

While static figures in textbooks and presentation slides have long been a mainstay of presenting concepts to undergraduates, interactive and dynamic versions are more interesting and arguably more effective. The pedagogical efficacy of interactive web-based Physlet tools, long used by physics educators, was described by Weiman et al (2008). Giving students the ability to interactively explore the impact of changing variables and parameters brings concepts to life and improves student engagement.

The ability of GAI to rapidly produce Colab notebooks with sophisticated interactive cells allows instructors to rapidly generate customized versions of interactive visualizations. An example is shown in Figure 9, from our `ClusterLRG_Rubin.ipynb` Colab notebook. The notebook cell generates a plot of flux measurements (astronomical magnitudes) vs. flux ratios (astronomical color) for galaxies around the Abell 360 cluster. The slider bars pick out an annular region between R and dR. This lets students explore the relative abundance of red elliptical cluster galaxies (with color (g-r) ~ 1) vs. field galaxies, and make a color-based photometric -redshift estimate.

We speculate that this may eventually lead to fully interactive, student-responsive GAI successors to our current textbooks. These tools could provide students with a rich and personalized learning experience that adapts to their learning style, with rapid iteration through the learn-challenge-deeper learning loop.

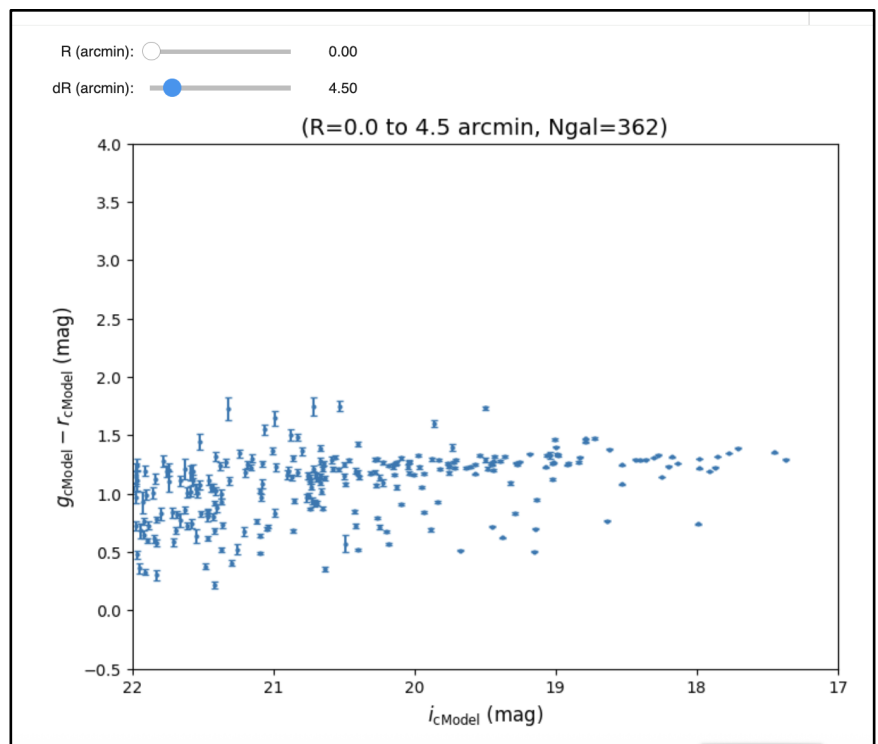

Figure 9. Generative AI allowed our instructional team to efficiently develop instructional Colab notebooks with interactive, dynamic figures such as this one. Rubin Observatory photometry (colors and magnitudes) of galaxies around the Abell 360 cluster from Data Preview 1 allows students to explore the radial dependence of "red sequence" elliptical galaxy colors in an annular region between R and dR, using the slider bars to set those parameters. This aids in visualizing the input data for the photometric redshift fit conducted by the cell shown in Figure 5.

## 4. Constructing and refining assignments and assessments

Each module in the course started with a substantial reading assignment in the course's textbook. To both incentivize students and to provide us with insight into their level of understanding, each reading assignment had an associated "reading quiz" that was administered online using the Canvas course management system. The production of reading quiz questions is, in our experience, a time-consuming endeavor. We used the NotebookLM utility to generate an initial bank of questions for each assigned chapter. A PDF copy of the textbook was uploaded to the NotebookLM tool as persistent context material. These draft questions were edited, expanded, and revised as appropriate before uploading into a Quiz in the Canvas system. Although we experimented with generating QTI zip files that allow for bulk ingestion of quiz questions, we found it less painful to cut-and-paste individual questions and answers one at a time. Streamlining this workflow is a good area for further development.

We used a similar GAI approach to generate a bank of draft multiple-choice questions for the final exam, including a practice exam used by the students as part of the review process. An extended bank of candidate short-answer questions for the final exam were constructed by the course instructional staff.

We found it particularly useful to submit successive draft exams to a browser-chat-interface GAI system, and have it "take" the exam. In numerous instances this identified poor or ambiguous wording, where the GAI misinterpreted the intent of the question. Questions would then be edited and resubmitted into a freshly-started chat, to eliminate stored context and history. Our draft exams went through multiple iterative quality-control evaluations by the course instructional staff.

Using GAI to generate and refine reading quizzes and to produce draft exam questions was a productivity boost- we generated these materials in less time than it would otherwise have taken. The real gain in course content came from using GAI for generating Colab notebooks at a level of sophistication that would have been otherwise unachievable, given the curricular development resources available to us.

## 5. Github repository generation

By the end of the Fall 2025 term, the Claude Code command line GAI tool was available, and we used that to clean up the course's Colab notebooks (we had used a variety of data access methods over the course of the term), test them, create a GitHub repository, and generate an instructor's guide to the notebooks and associated data sets. The initial prompt for this was:

> *First, review each notebook and make sure all the files that are needed are included in this directory. Tell me about any that are missing. Next, we need to clean up the ways the files are accessed. We should edit the notebooks to assume the files needed are in the same working directory as the one from which the notebooks are being run. start by making a subdirectory called 'originals' and make backup copies of all ipynb files there before editing them. then edit the files as needed to access data files assuming they are local. Change the comments as needed to do this. Next, we need to test the notebooks, one at a time, and make sure they all work. Once that is verified, we need to produce an instructor's guide that describes each notebook in turn, along with learning goals, pip install prerequisites, and other useful stuff.*

> *This is all material from Fall 2025 version of an introductory extragalactic astronomy and cosmology course at Harvard. Go ahead and put me down as the technical contact person. Be sure to add an acknowledgements section to which I'll add names. I'll need the instructor guide in Latex to upload to Overleaf.*

This entire operation, with a few iterations, took less than one hour to complete. Minor edits were made to the Latex instructor guide, which was then added to the GitHub repository. Without the added efficiency that GAI provided, it is extremely unlikely we would have invested the effort needed to make the Colab notebooks usefully available. The GitHub repository contains a dozen high-end Colab notebook/worksheets that were generated for this course. Some lay foundational skills in diffraction grating physics and data analysis, others bear upon observational extragalactic astronomy, and they culminate in an exercise that uses supernova luminosity distances data vs. redshift to have students perform Hubble diagram fits that indicate accelerating cosmic expansion.

**LESSONS LEARNED**

We re-iterate that the course enrollment of 14, and the even more limited survey participation, clearly puts us in the regime of small-number statistics. Broadly speaking, we consider the course to have been a success. Student performance on the final exam indicated strong progress towards the learning goals. Student satisfaction ratings for the course were not statistically different from the prior year, despite an expanded set of learning goals. In the course evaluation about half the students chose either "strongly agree" or "agree" with the statement that the GAI aspects enhanced their learning.

Arguably the largest strength of GAI use was the ease of generation and debugging of sophisticated Colab notebooks by both the instructional staff and, by the end of the term, the students as well. Course administration and the production of exams and assignments was more time efficient.

Two aspects of this experience came as a surprise to us. The first was the lack of student interest in using the GAI+Slack channel. We had hoped to establish a community spirit of shared engagement, by expanding on the 1:1 GAI chat architecture. We have heard that the Harvard Business School established a similar GAI+Slack system, except they allowed private Direct Messaging between students and the AI entity, which ours (intentionally) did not. Students at the Business School evidently much preferred to ask questions of clarification over a private GAI channel rather than in a shared public space. We speculate that the students in our course had similar reluctance. The second surprise was that second year undergraduates did not share our sense that GAI is an increasingly integral element of modern science. Their sympathy for this aspect of the course increased after we invested substantial classroom time in explaining the rationale. It would have been far better to do this during the first week of class.

Suggestions for GAI developers that would enhance its application to situations like this one include:
- Improved integration with overall workflow, such as moving quiz questions into Canvas,
- Provide to access to use statistics (anonymized would be fine) for NotebookLM and GEM use.
- Increased file upload capacity.

Stepping back to view this course through a broader lens, students analyzed observational data from WMAP, SDSS, DESI, and Rubin using Colab+GAI notebooks. This empowered students to reproduce some of the classic pillars of observational cosmology. In parallel the students enhanced their skills in data analysis methodology, augmented by generative artificial intelligence.

**NOTEBOOK AVAILABILITY**

The Colab notebooks, associated data files, and a terse instructor's guide are available at https://github.com/astrostubbs/harvard-astro17-labs.

**ACKNOWLEDGEMENTS**

We are grateful to Google for making the Gemini GAI suite of tools readily available to our community, and to our colleagues in Harvard University IT for their system-level support. Heartfelt thanks to the Fall 2025 cohort of Astronomy 17 students for their patience and participation in this course. Open access to data from the WMAP, SDSS, and Rubin projects allowed our students to undertake the hands-on learning projects that were essential to the course. The manuscript was written by the authors, without AI tools.